# Spatiotemporal Trajectory Tracking Method for Vehicles Incorporating Lead-Lag Judgement

Yuan Li[a,1], Xiang Dong[a,1], Tao Li[a,1], Junfeng Hao[b], Xiaoxue Xu[a], Sana Ullah[a], Yincai Cai[a], Peng Wu[a], Ting Peng[a,*]

[a]*Key Laboratory for Special Area Highway Engineering of Ministry of Education, Chang'an University, Xi'an, China*

[b] *China Railway No.7 Bureau Group No.3 Engineering Co.,Ltd. Xi'an*

**Abstract**

In the domain of intelligent transportation systems, especially within the context of autonomous vehicle control, the preemptive holistic collaborative system has been presented as a promising solution to bring a remarkable enhancement in traffic efficiency and a substantial reduction in the accident rate, demonstrating a great potential of development. In order to ensure this system operates as intended, accurate tracking of the spatiotemporal trajectory is of crucial significance. Moreover, minimizing the tracking error is a necessary step in this process. To this end, a novel lead-lag judgment mechanism is proposed. This mechanism precisely quantifies the longitudinal positional deviation between the vehicle and the target trajectory over time, then the deviation is corrected with a real - time acceleration compensation strategy, as a result, the accuracy and reliability of trajectory tracking are significantly enhanced. Real - vehicle experiments were conducted in a dedicated test field to validate the feasibility of this innovative approach empirically. Subsequently, the obtained tracking data was subsequent processed using the lead-lag judgment mechanism. In this step, we carefully analyzed the spatiotemporal error patterns between the vehicle and the target trajectory under different alignments and speeds. Finally, using real highway speed and alignment data, we conducted comprehensive spatiotemporal trajectory tracking simulations. Through experiments and simulations, tracking errors maintained in an acceptable range and reasonable spatiotemporal distance is given during the preemptive merging process on highway ramps. Overall, this study offers valuable insights for highway ramp emerging safety. Future work can expand on these findings.

*Key words*: Preemptive holistic collaborative system; Vehicle spatiotemporal trajectory tracking; Integrated horizontal-vertical control algorithm; Traffic capacity; System performance optimization

## 1. Introduction

Vehicle trajectory tracking control has become a focal point in automotive engineering research, primarily categorized into two types: algorithms based on geometric kinematic models and those based on vehicle kinematic and dynamic models. The former includes the Pure Pursuit algorithm and Stanley algorithm. Pure Pursuit is robust but highly dependent on lookahead point selection and struggles to achieve optimality. Improvements have focused on optimizing lookahead distance, such as the approach by Horváth et al. [1] and the model predictive control (MPC)-based method by Kim et al. [2]. The Stanley algorithm excels in low-speed stability and small tracking errors but cannot balance precision and smoothness. Recent advancements involve parameter optimization using genetic algorithms [3,4], integration with other algorithms [5], and the incorporation of discrete predictive models [6,7].

Algorithms based on vehicle kinematic and dynamic models encompass a variety of control strategies, including proportional-integral-derivative (PID) control, linear quadratic regulator (LQR), MPC, sliding mode control (SMC), and active disturbance rejection control (ADRC). PID control, a classic feedback method, is widely used for its simplicity and adaptability, with enhancements achieved through parameter tuning and hybrid control approaches [8-10]. LQR, suitable for linear dynamic systems, has been improved by adding feedforward control, particle swarm optimization, and fuzzy control, or combined with other algorithms for better performance [11-14].

---

*Correspondence author. E-Mail: t.peng@ieee.org

[1]These authors contributed equally to this work.

MPC, despite its high precision and robustness, faces challenges in real-time implementation due to its computational complexity, often necessitating integration with other methods [15]. SMC, while effective in handling vehicle dynamics, can cause chattering and has lower control accuracy, which has been addressed through various modifications [16,17]. ADRC, known for its strong disturbance rejection capability, requires precise disturbance modeling and parameter tuning, as demonstrated by Kang et al. [18].

Despite these advancements, existing research predominantly focuses on spatial trajectory tracking, neglecting the temporal discrepancies between actual and ideal trajectories caused by factors such as road conditions, vehicle performance, and external environments. These discrepancies can significantly impact vehicle operation assessment and control decisions. In the context of highway ramp merging zones, a critical bottleneck in traffic flow, intelligent connected autonomous vehicles (CAVs) offer a promising solution through real-time communication and precise trajectory control. Current research on CAVs' cooperative merging strategies at ramps mainly adopts rule-based, optimization-based, and learning-based approaches. Rule-based methods, such as the one proposed by Ding et al. [19], provide near-optimal merging sequences with low computational costs. Optimization-based methods, including mixed-integer nonlinear programming [20,21], game theory [22,23], and hierarchical control [24-26], aim to determine optimal merging strategies from a global perspective. Learning-based methods, particularly reinforcement learning [27-29], show potential in achieving human-like intelligent decision-making for optimal control strategies.

Previous studies primarily concentrated on spatial trajectory tracking, neglecting to analyze traffic from a holistic perspective [30,31] and overlooking the disparities between actual and ideal trajectories in the temporal dimension. These differences, which are caused by factors such as traffic conditions, vehicle power systems and sudden environmental changes, lead to deviations in the time dimension. These deviations are significant because they have an impact on vehicle operation evaluation and control decision-making, and may even pose risks to safety. Peng et al. [30],[32] proposed a distributed real-time information-sharing mechanism based on section management units, thereby real-time sharing of vehicle and infrastructure status information as well as driving intentions within a local area to enhance traffic safety and efficiency, preemptive holistic collaborative road transportation system is constructed. Building on this, Peng et al. [33] developed a cooperative control strategy prioritizing either the mainline or the ramp, predefining vehicle trajectories through the information-sharing mechanism. Hence, preemptive Spatiotemporal trajectories can solve this problem and improve tracking efficiency.

To surmount this limitation, our study adopts a horizontal and vertical decoupling strategy. Specifically, the LQR algorithm is employed for lateral control, and the dual PID algorithm is utilized for longitudinal control. Moreover, a temporal factor is incorporated through the introduction of an advance/delay judgment mechanism and a real-time acceleration compensation strategy. This comprehensive approach is designed to achieve Spatiotemporal trajectory tracking and to augment the efficiency and precision of vehicle control.

## 2. Spatial-temporal Trajectory Tracking

During the actual driving process, influenced by multiple factors such as road conditions, vehicle performance, and the external environment, the actual vehicle trajectory often exhibits a longitudinal position deviation in the time dimension, either ahead of or behind the ideal trajectory. The accumulation of these deviations may significantly affect the evaluation of vehicle operation and control decisions.

To achieve precise Spatiotemporal trajectory tracking, this paper proposes an advance - delay determination mechanism. This mechanism is used to quantify the longitudinal position deviation of the actual vehicle trajectory relative to the ideal trajectory in the time dimension. It is then combined with a real - time acceleration compensation strategy for correction. Specifically, when the vehicle's longitudinal position is ahead of the target point, negative acceleration compensation is applied; when the vehicle's longitudinal position lags behind the target point, positive acceleration compensation is applied, enabling the vehicle to gradually approach the ideal trajectory. This method fully takes into account the dynamic characteristics of the vehicle. By means of real - time deviation correction, it aims to enhance the accuracy and stability of trajectory tracking.

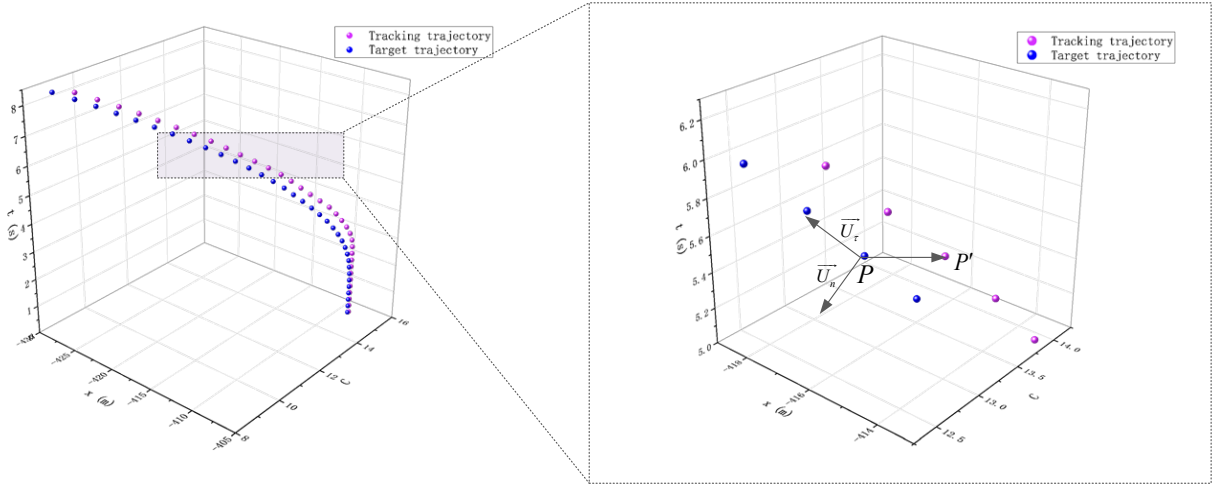

Fig. 1 Judgment Diagram

## 2.1. Judgment Criterion

Consider the dot-product of the vector formed by the target trajectory point and the tracking trajectory point at the same moment with the unit vector in the forward direction. This dot-product is used to determine whether the tracking trajectory point is ahead of or behind the target trajectory point. The result of the dot-product represents the offset of the tracking point relative to the target point in the forward direction.

During $0 \leq t \leq T$, the target trajectory point is denoted as $P: x(t)\vec{i} + y(t)\vec{j}$, and the tracking trajectory point is denoted as $P': x'(t)\vec{i} + y'(t)\vec{j}$, at any moment $t_0$, there exists:

Calculate the vector (spatial position difference) formed by the target trajectory point and the tracking trajectory point at time $t_0$:

$$\overrightarrow{PP'}_{t_0} = [x'(t_0) - x(t_0)]\vec{i} + [y'(t_0) - y(t_0)]\vec{j} \quad (1)$$

Calculate the vector formed by the target trajectory point at time $t_0$ and the target trajectory point at time $t_0 + \Delta t$, where $\Delta t \to 0$. This computation essentially determines the instantaneous velocity of the target trajectory point at time $t_0$.

$$\vec{v}_{t_0} = \lim_{\Delta t \to 0} \left[ \frac{x(t_0 + \Delta t) - x(t_0)}{\Delta t}\vec{i} + \frac{y(t_0 + \Delta t) - y(t_0)}{\Delta t}\vec{j} \right] \quad (2)$$

Normalize this vector as shown in Equation (3), with its direction representing the tangent at the target trajectory point on the experimental orbit, pointing in the direction of the advancing target trajectory point.

$$\overrightarrow{U_\tau} = \frac{\lim_{\tau \to 0}\left[\frac{x(t_0+\Delta t)-x(t_0)}{\Delta t}\vec{i} + \frac{y(t_0+\Delta t)-y(t_0)}{\Delta t}\vec{j}\right]}{\sqrt{\left(\frac{x(t_0+\Delta t)-x(t_0)}{\Delta t}\right)^2 + \left(\frac{y(t_0+\Delta t)-y(t_0)}{\Delta t}\right)^2}}$$
$$= U_{\tau x}\vec{i} + U_{\tau y}\vec{j} \quad (3)$$

$$\begin{cases} \overrightarrow{PP'} \cdot \overrightarrow{U_\tau} > 0, & \text{lead} \\ \overrightarrow{PP'} \cdot \overrightarrow{U_\tau} < 0, & \text{lag} \end{cases} \quad (4)$$

At time $t_0$, calculate the unit vector $\overrightarrow{U_n}$ perpendicular to the direction of advancement (as shown in Equation (5)). Then, perform a dot product between this unit vector U and the vector formed by the target trajectory point and the tracking trajectory point (with the direction pointing from the target trajectory point to the tracking trajectory point). This dot product result serves as an indicator of the lateral deviation of the tracking point relative to the target point, determining whether the tracking trajectory point is to the left or right of the target trajectory point.

$$\overrightarrow{U_n} = -U_{\tau y}\vec{i} + U_{\tau x}\vec{j} \quad (5)$$

$$\begin{cases} \overrightarrow{PP'} \cdot \overrightarrow{U_n} > 0, & \text{veer left} \\ \overrightarrow{PP'} \cdot \overrightarrow{U_n} < 0, & \text{veer right} \end{cases} \quad (6)$$

When both $\overrightarrow{PP'} \cdot \overrightarrow{U_\tau} = 0$ and $\overrightarrow{PP'} \cdot \overrightarrow{U_n} = 0$, it signifies that the position of the tracking trajectory point coincides with that of the target trajectory point.

*2.2. Acceleration Correction*

To prevent frequent adjustments due to minor errors that may lead to vehicle instability, this paper establishes a longitudinal error threshold of $\Delta p_{Threshold} = \pm 0.5m$. Interventions are only implemented when the error exceeds this threshold, ensuring smooth vehicle operation. No acceleration compensation is required when condition $|\Delta p(t)| = |\overrightarrow{PP'} \cdot \overrightarrow{U_\tau}| \leq 0.5$ is met; however, when condition $|\Delta p(t)| = |\overrightarrow{PP'} \cdot \overrightarrow{U_\tau}| > 0.5$ is satisfied, acceleration compensation $a_{Compensation}(t)$ must be introduced to counteract longitudinal errors that exceed the threshold range.

$$a_{Compensation}(t) = \begin{cases} -\dfrac{2(\Delta p(t) - 0.5)}{T_\omega^2}, & \Delta p(t) > 0.5 \text{ (lead)} \\ -\dfrac{2(\Delta p(t) + 0.5)}{T_\omega^2}, & \Delta p(t) < -0.5 \text{ (lag)} \end{cases}$$
(7)

The core idea of real-time acceleration compensation is to dynamically adjust the acceleration based on the longitudinal position difference (lead or lag) between the vehicle and the target trajectory point at the same moment during tracking, ensuring that the vehicle can reasonably approach or maintain the progress of the target trajectory.

**3. Real-world vehicle experiment**

A test platform for real-world vehicle experiments has been established by programming in C++ to develop the required algorithm packages based on the Robot Operating System (ROS). Subsequently, other environmental dependencies are deployed onto an industrial computer to complete the construction of the test platform.

This experiment comprises two distinct parts. Initially, we employed software tailored for this experiment to gather trajectory data from three segments: a 200-meter straight track, a 200-meter curved track, and a 400-meter combined straight and curved track. The trajectory data was collected at a resolution of one data point per meter, encompassing details such as the vehicle's position, speed, heading angle, and other pertinent information corresponding to various time points. In the subsequent part, we utilized an autonomous driving control algorithm (LQR+dual PID) to track the previously collected trajectories. Each trajectory segment was tested three times, and the actual tracking trajectories obtained during each test were collected and compared with the trajectories collected in the first part to ascertain the tracking errors.

Experimental data is quantitatively assessed based on four primary error indicators: speed error, heading angle error, lateral error (left/right deviation error), and advance/delay error. Specifically, the speed error is utilized to measure the velocity deviation between the target trajectory and the tracked trajectory at each time point. The heading angle error reflects the discrepancy in heading angles between the target trajectory and the tracked trajectory at each time point. The analysis of lateral error and advance/delay error, on the other hand, is conducted through quantitative evaluation of post-processed real-vehicle experimental data, based on the judgment criterion proposed in Section 2.1.

In straight segment tests, vehicles were tested at speeds of 20, 25, and 30 km/h. Speed mainly affected longitudinal Spatiotemporal relationships, with limited indirect impact on path deviation and heading control. As speed increased, dynamic adjustment time for speed error grew, and fluctuations during acceleration/deceleration were more pronounced. However, heading angle and lateral errors changed little.

In curved segment tests at 10, 12, and 15 km/h, errors generally increased with speed, including expanded speed error range, increased negative heading angle error, and larger lateral rightward deviation. Errors were interrelated, with speed error influencing steering inertia, which affected heading angle error, which in turn caused lateral error changes. Speed error was also directly related to Spatiotemporal relationships, jointly affecting trajectory accuracy.

For the purpose of conducting a more in-depth analysis of the Spatiotemporal discrepancies between the target trajectory and the tracked trajectory, we have devised an experimental scheme that integrates both linear and curved segments. The detailed experimental protocol is outlined in Table 1.

Table. 1 Testing Scheme for Driving on Combined Straight and Curved Road Segments

| Number | Straight Speed | Curved Speed |
|---|---|---|
| 1 | 20km/h | 10km/h |
| 2 | 25km/h | 12km/h |
| 3 | 30km/h | 15km/h |

When the test speed is set to 20 km/h for the straight-line segment and 10 km/h for the curved segment, the results of various error indicators are presented in Figure 2. During the initial approximately 10 seconds of the straight-line segment test, the speed error fluctuates significantly within an interval denoted as m. In the later stage of the straight-line segment test, the speed error primarily fluctuates around ±0.4 m/s. Upon entering the curved segment, the magnitude of the speed error decreases, primarily distributed within ±0.1 m/s.

When the test speeds are elevated to 25 km/h for the straight segment and 12 km/h for the curved segment, the results of various error indicators are depicted in Figure 3. Initially, the speed error fluctuates prominently within the first few seconds of the straight segment, but gradually stabilizes later. Upon entering the curved segment, the speed error diminishes. The heading angle error converges rapidly but fluctuates more significantly in the curved segment.

When further increasing the test speed to 30 km/h on straight sections and 15 km/h on curved sections, the results for various error indicators are shown in Figure 4. In terms of speed error, significant fluctuations, predominantly negative, were observed within the first approximately 16 seconds on straight sections, with a fluctuation range of approximately m/s. Subsequently, the speed error gradually stabilized, narrowing the fluctuation range to m/s. Upon entering curved sections, the speed error further decreased, primarily staying within m/s. Regarding heading angle error, it rapidly converged from an initial value of approximately -0.03 rad to near zero within the initial stage on straight sections, later fluctuating slightly within ±0.005 rad. Upon entering curved sections, due to increased steering demands, the heading angle error fluctuations intensified, predominantly positive, reaching a maximum error of 0.067 rad.

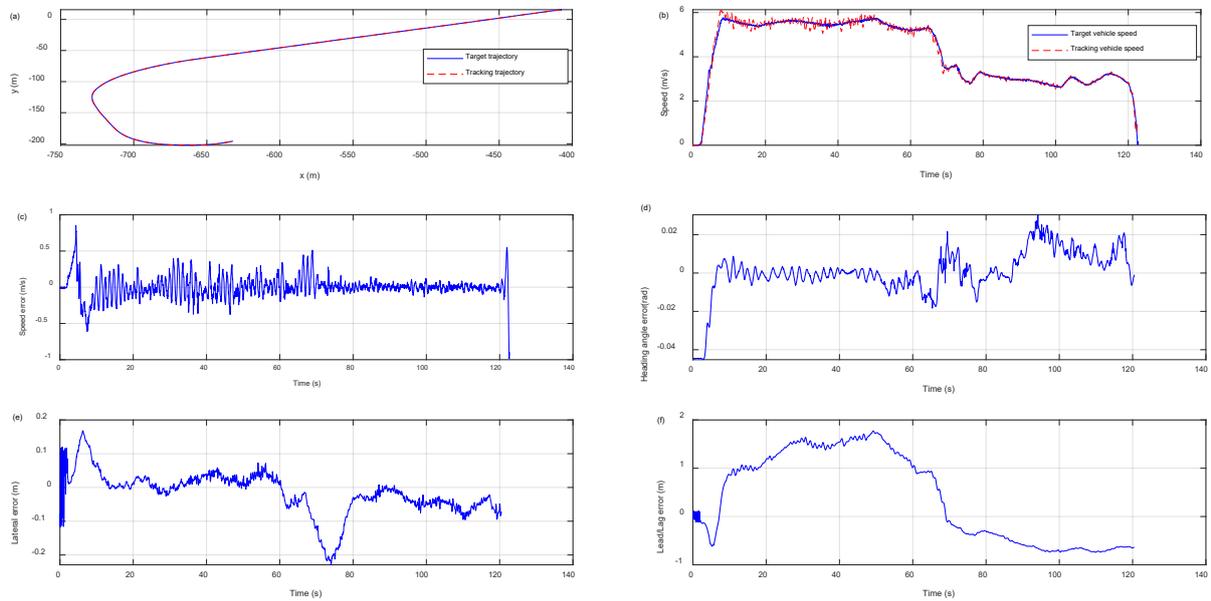

Fig. 2 Test Results for the Combined Road Segment (mix2010) (a)Driving Path. (b)Speed Comparison Analysis. (c)Speed Error Distribution. (d)Heading Angle Error. (e)Lateral Position Error. (f) Lead/Lag Time Error.

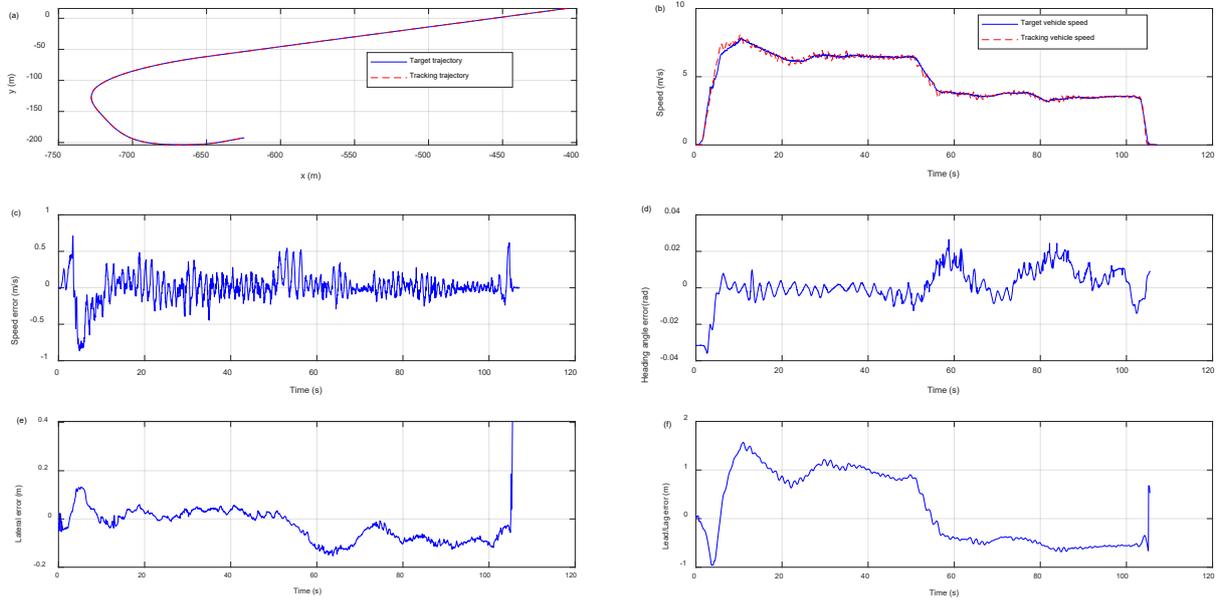

Fig. 3 Test Results for the Combined Road Segment (mix2512) (a)Driving Path. (b)Speed Comparison Analysis. (c)Speed Error Distribution. (d)Heading Angle Error. (e)Lateral Position Error. (f) Lead/Lag Time Error.

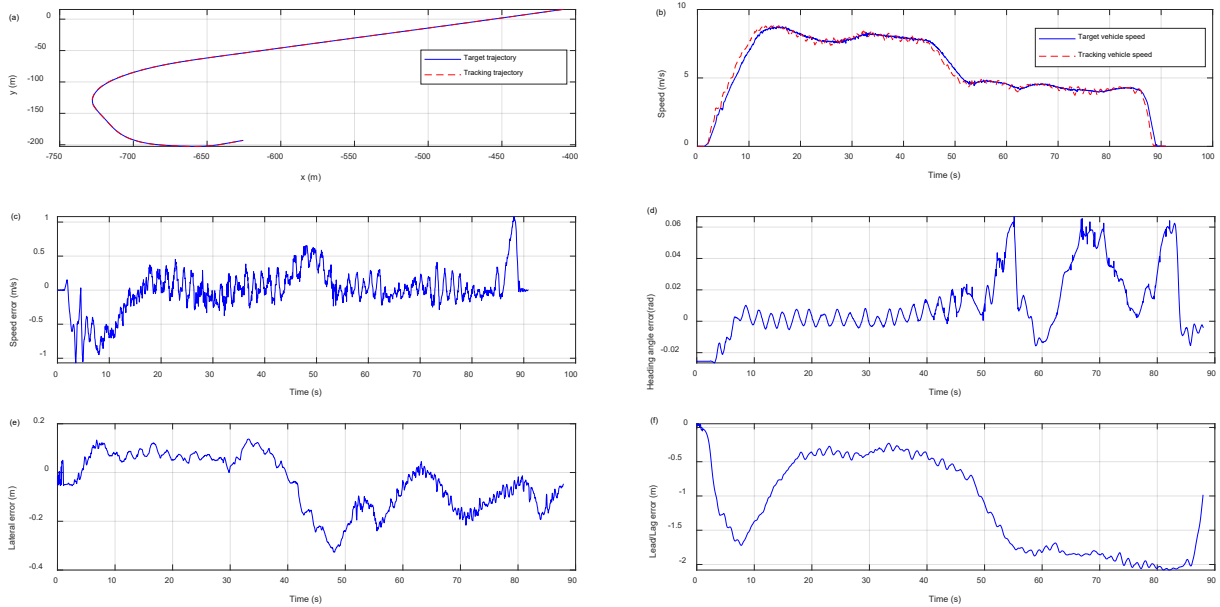

Fig. 4 Test Results for the Combined Road Segment (mix3015) (a)Driving Path. (b)Speed Comparison Analysis. (c)Speed Error Distribution. (d)Heading Angle Error. (e)Lateral Position Error. (f) Lead/Lag Time Error.

In the combined road segment scenario, errors generally increase with rising speeds. Speed errors fluctuate more during left turns, with absolute values and frequencies escalating. Heading angle and lateral errors also broaden with increasing speed during these turns. In Spatiotemporal terms, delay errors appear and accumulate sooner at higher speeds, influenced by factors such as vehicle dynamics and information processing delays. This is evident in the mix3015 scenario, highlighting increased matching difficulty between vehicle position and target trajectory under complex driving conditions and speed variations, thereby underscoring the detrimental effect of speed on trajectory accuracy.

## 4. Simulation Experiment

This study is grounded in the original data including straight lines, curves, and turning angle tables. Employing the HintCAD road design software, the planar alignment diagram of the TJ1 contract section of the Huili - Luquan (Sichuan section) Expressway on S81 line has been meticulously completed. With a comprehensive consideration of the path alignment characteristics and simulation calculation efficiency, the section ranging from K0+000.00 to K1+594.433 was selected, and the center - pile coordinates were exported at intervals of 1-meter for subsequent simulation-based experimental research.

Data processing and algorithm implementation were carried out by relying on Python software. Its core procedures are as follows: Initially, the original center - pile data with a 1 meter collection interval, generated by the HintCAD software, is imported to construct the path diagram, which serves as a fundamental basis for subsequent analysis. Regarding the setting of the vehicle driving speed, the vehicle first accelerates to the desired velocity of 100 km/h, then sustains a constant-speed driving state, and gradually decelerates as the vehicle approaches the end of the path. Taking into account the speed fluctuations of vehicles in real - world scenarios, a 2% random speed error was incorporated, and the vehicle speed and heading angle were adaptively adjusted in accordance with the curvature characteristics of the original path. After the aforementioned processing steps, the target trajectory data, encompassing time - related information and key parameters such as X-coordinate, Y-coordinate, speed, and heading angle at the corresponding time, was ultimately generated.

The core essence lies in that, based on the target speed and heading angle, an advance-delay judgment mechanism is introduced to quantitatively assess the longitudinal position deviation between the vehicle tracking trajectory and the target trajectory in the time dimension, and it is rectified in combination with the real-time acceleration compensation strategy. Specifically, when the longitudinal position of the vehicle is ahead of the target point, negative acceleration compensation is imposed; when the longitudinal position of the vehicle lags behind the target point, positive acceleration compensation is applied to enable the vehicle to gradually approach the target trajectory.

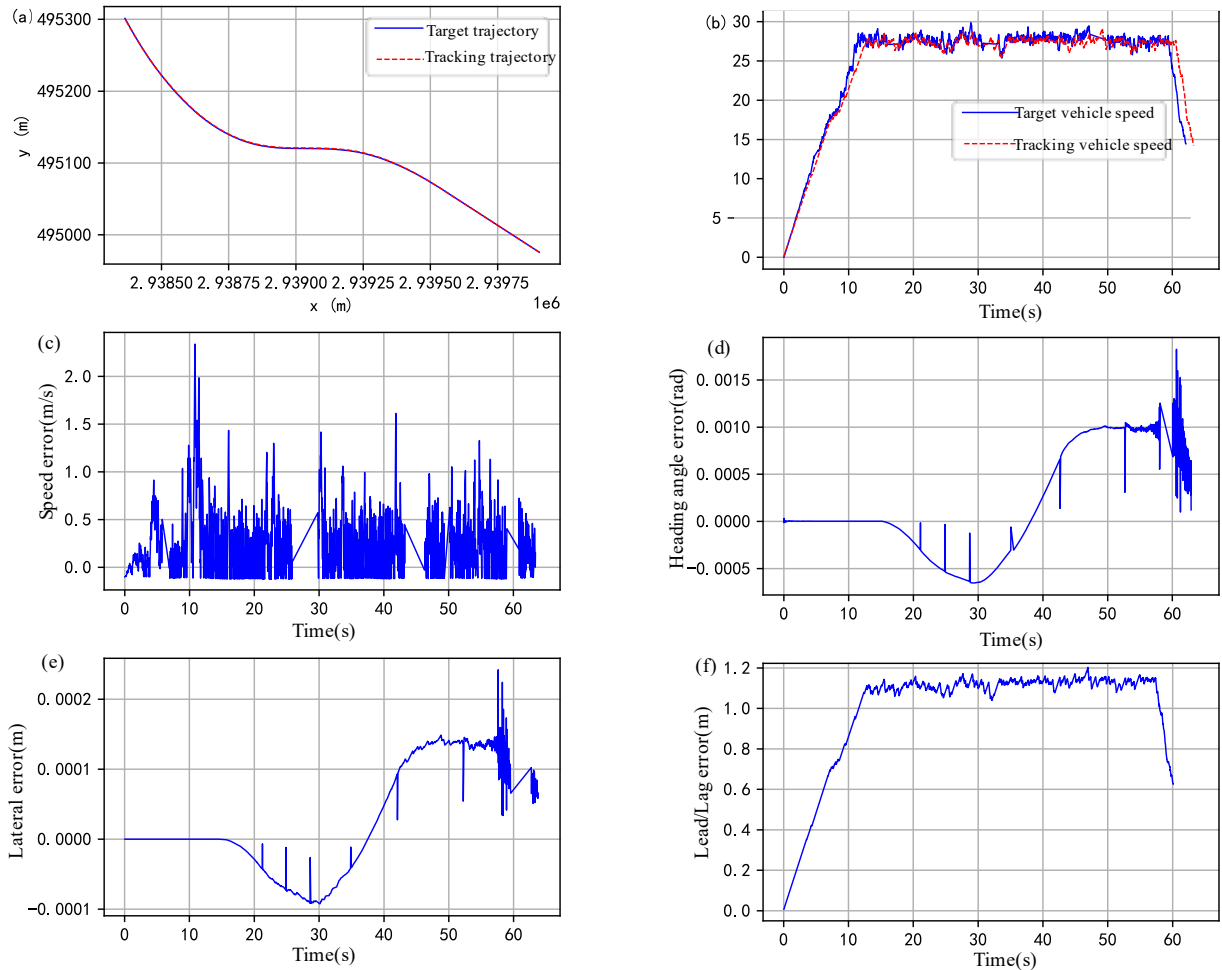

Fig. 5 Test Results of Simulation Experiment (a)Driving Path. (b)Speed Comparison Analysis. (c)Speed Error Distribution. (d)Heading Angle Error. (e)Lateral Position Error. (f) Lead/Lag Time Error

The simulation results are illustrated in Figure 5 with the initial coordinate located at the bottom right of the driving path map and a positive heading angle assumed. According to the results, speed errors mainly fluctuate within [-0.1, 0.5] m/s, heading angle errors are generally small, within [-0.0005, 0.0015] rad, and lateral errors are also minimal, within [-0.0001, 0.00026] m. Combining Figures 5c) and 5f), despite the tracking vehicle's instantaneous speed generally being lower than the target speed, due to initial position, dynamic adjustment mechanisms driven by lead/lag values, and nonlinear speed-position relationships, its spatial position exhibits an advance phenomenon. Specifically, within the first 3.6 seconds of the simulation, the tracking vehicle's instantaneous speed was higher than the target speed for most of the time, leading to a gradual accumulation of early errors up to 0.5 m, triggering the system to correct by reducing acceleration. This adjustment mechanism somewhat widened the instantaneous speed difference, causing the tracking vehicle's instantaneous speed to further drop below the target speed at some points. However, due to the small speed difference and short accumulation time, the tracking vehicle's spatial position may still advance relative to the target trajectory point during the simulation. As shown in Figure 5(f), the maximum Spatiotemporal position error throughout the process is approximately 1.21 m.

To intuitively understand the spatial position matching characteristics between tracking points and target points at various times, this paper extracts target trajectory and tracking trajectory data from the first 10 seconds and the last 4 seconds of the simulation process and plots their Spatiotemporal trajectories. According to Figure 6(a), within the first 10 seconds, the tracking trajectory points at each moment precede the target trajectory points, with this advancement error accumulating over time. Figure 6(b) indicates that, although the tracking trajectory points at each moment still precede the target trajectory points, the advancement trend gradually diminishes.

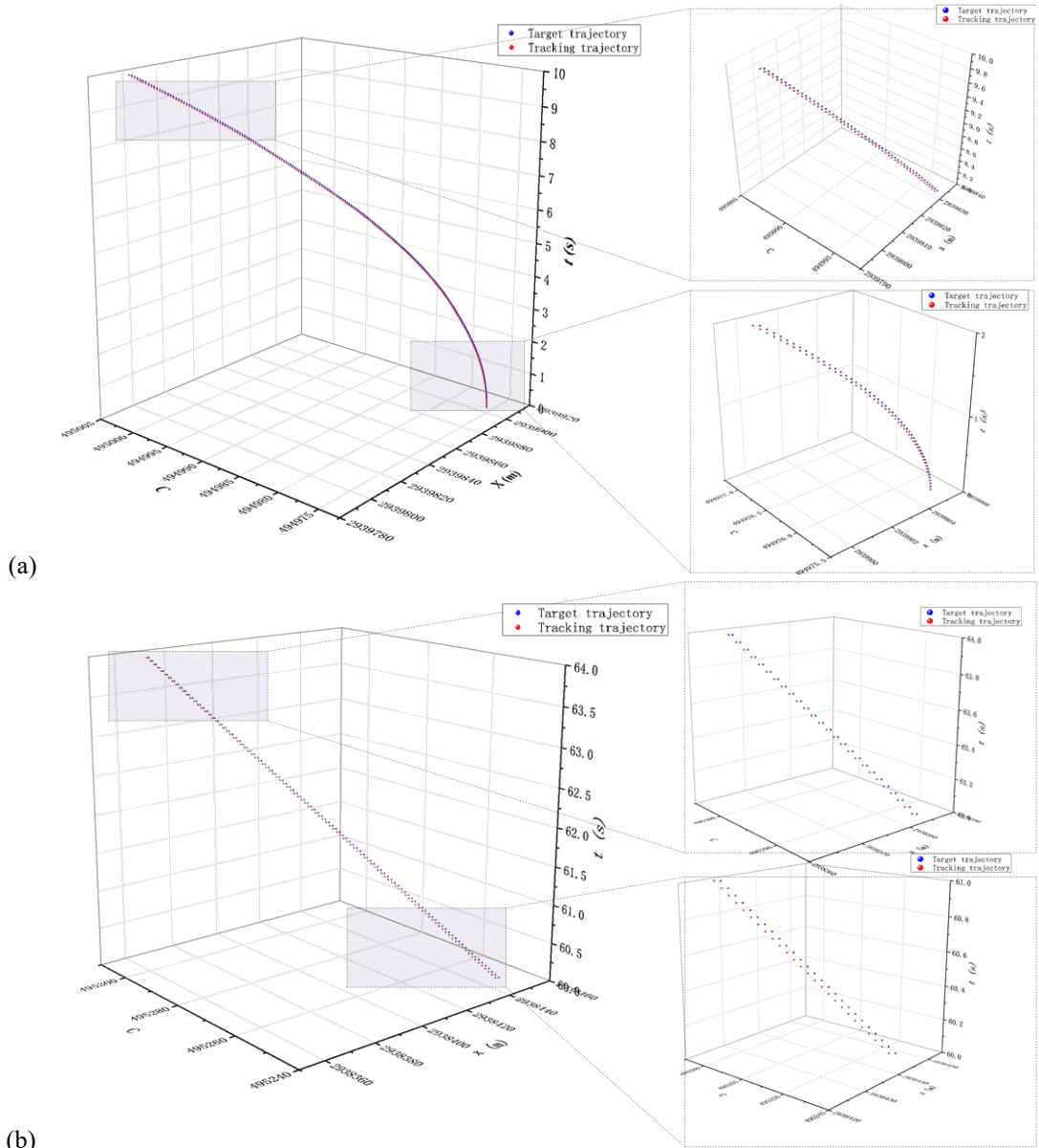

**Fig. 6** Spatiotemporal diagram of target trajectory and tracking trajectory. (a) the First 10 Seconds of the Route. (b) the Last 4 Seconds of the Route.

## 5. Conclusion

In this paper, we explore the precise control and error analysis of vehicle spatiotemporal trajectory tracking in complex road scenarios. A control algorithm is designed and validated to establish safe spatiotemporal distances for highway ramp merging. Key findings include:

(1) Real-vehicle experiments using LQR and dual PID algorithms were conducted across different road alignments and speeds, with an lead/lag judgment criterion for data post-processing. Results showed increasing spatiotemporal errors with higher speeds and more complex alignments, peaking at about 2 meters in combined straight-curve sections.

(2) An lead/lag judgment mechanism, coupled with real-time acceleration compensation, was proposed and tested in simulations based on actual highway conditions. At a desired speed of 100 km/h, the algorithm achieved a spatiotemporal error of approximately 1.21 meters, validating its effectiveness.